\documentclass[runningheads]{llncs}

\usepackage[utf8]{inputenc} % allow utf-8 input
\usepackage[T1]{fontenc}    % use 8-bit T1 fonts
\usepackage{hyperref}       % hyperlinks
\usepackage{url}            % simple URL typesetting
\usepackage{booktabs}       % professional-quality tables
\usepackage[dvipsnames, table]{xcolor}
\usepackage{amsfonts,amsmath}       % blackboard math symbols
\usepackage{nicefrac}       % compact symbols for 1/2, etc.
\usepackage{microtype}      % microtypography
\usepackage{graphicx}
\usepackage[sorting=none, backend=bibtex]{biblatex}
\usepackage[capitalise]{cleveref}
\usepackage{todonotes}
\usepackage{siunitx}
\usepackage{amsfonts, amsmath, amssymb}
\usepackage{indentfirst}
\usepackage{pifont}% http://ctan.org/pkg/pifont
\bibliography{references}

\usepackage{hyperref}       % hyperlinks
\usepackage{xcolor}
\usepackage{sidecap}
\hypersetup{
    colorlinks,
    linkcolor={blue!50!black},
    citecolor={blue!50!black},
    urlcolor={blue!50!black}
}
\begin{document}
\title{Analysing race and sex bias in brain age prediction}
\author{Carolina Piçarra and Ben Glocker}
% index{Piçarra, Carolina}
% index{Glocker, Ben}
%
%\authorrunning{B. Glocker et al.}
\institute{Department of Computing, Imperial College London, UK\\\email{c.picarra@imperial.ac.uk}}
\maketitle              % typeset the header of the contribution
\begin{abstract}

Brain age prediction from MRI has become a popular imaging biomarker associated with a wide range of neuropathologies. The datasets used for training, however, are often skewed and imbalanced regarding demographics, potentially making brain age prediction models susceptible to bias. We analyse the commonly used ResNet-34 model by conducting a comprehensive subgroup performance analysis and feature inspection. The model is trained on 1,215 T1-weighted MRI scans from Cam-CAN and IXI, and tested on UK Biobank (n=42,786), split into six racial and biological sex subgroups. With the objective of comparing the performance between subgroups, measured by the absolute prediction error, we use a Kruskal-Wallis test followed by two post-hoc Conover-Iman tests to inspect bias across race and biological sex. To examine biases in the generated features, we use PCA for dimensionality reduction and employ two-sample Kolmogorov-Smirnov tests to identify distribution shifts among subgroups. Our results reveal statistically significant differences in predictive performance between Black and White, Black and Asian, and male and female subjects. Seven out of twelve pairwise comparisons show statistically significant differences in the feature distributions. Our findings call for further analysis of brain age prediction models.

\end{abstract}

\section{Introduction}
The global population growth and longer life expectancy are linked to the rising prevalence of age-related neurodegenerative and neuropsychiatric diseases \cite{hou2019ageing, deuschl2020burden, dumurgier2020epidemiology}. As a result, there is an increasing need to establish connections between brain ageing and disease processes, to better understand their mechanisms and enable early detection and diagnosis. Significant research efforts have focused on investigating the potential of brain-predicted age as an indicator of how an individual's brain health may deviate from the norm \cite{baecker2021machine, cole2017predicting}. As a neuroimaging-driven biomarker, it has the potential of containing a broad spectrum of brain characteristics in a single measurement \cite{de2022mind}. Several studies have proposed brain age prediction for the characterisation of neuropathology \cite{cole2020longitudinal, rokicki2021multimodal}, epilepsy \cite{sone2021neuroimaging}, as well as an indicator of clinical risk factors \cite{cole2018brain, beck2022cardiometabolic}. Most studies used structural MRI, due to its common use in clinical settings and high resolution, capturing even small structural variations in brain anatomy. Deep learning (DL), and in particular convolutional neural networks (CNNs), are widely used models for brain age prediction from MRI \cite{tanveer2023deep,peng2021accurate}. Studies rely on well-established datasets, including the UK Biobank \cite{sudlow2015uk}, the Cambridge Centre for Ageing Neuroscience (Cam-CAN) dataset \cite{taylor2017cambridge}, IXI \cite{ixi}, the Alzheimer’s Neuroimaging Initiative (ADNI) dataset \cite{weiner2017recent}, the The Open Access Series of Imaging Studies (OASIS) \cite{marcus2007open}, among others. These datasets tend to be skewed and biased regarding ethnic and racial diversity, with a majority of White subjects. When models are trained on data with unbalanced demographics, the performance may degrade in relevant subgroups \cite{castro2020causality}. Thus, it is important to test such models for potentially disparate performance across subgroups. In this study, we analyse a ResNet-34 brain age prediction model by conducting a comprehensive statistical subgroup performance analysis and feature inspection.

\section{Materials and methods}

\paragraph{\textbf{Datasets}}

For training and validation of the brain age prediction model, we used the Cam-CAN \cite{taylor2017cambridge} and the IXI dataset with healthy volunteers. For testing, the UK Biobank dataset was selected due to its size and availability of race and biological sex information. The demographics for each dataset are available in Table \ref{demographic_table}. Patient racial information is not provided for the Cam-CAN dataset. However, considering that the data collection took place in Cambridge (United Kingdom), we assume the majority of volunteers were White. All scans from Cam-CAN and IXI were pre-processed by us using the following steps: 1) Lossless image reorientation using the direction information from the image header; 2) Skull stripping with ROBEX v1.2\footnote{\url{https://www.nitrc.org/projects/robex}} \cite{iglesias2011robust}; 3) Intensity-based rigid registration to MNI atlas ICBM 152 2009a Nonlinear Symmetric\footnote{\url{http://nist.mni.mcgill.ca/?p=904}}; 4) Bias field correction with N4ITK\footnote{\url{https://itk.org}} \cite{tustison2010n4itk}. The UK Biobank images were already skull-stripped and bias field corrected, and only the registration to MNI space was performed by us.

\begin{table}[]
\centering
\caption{Demographic information of all datasets used.}
\begin{tabular}{@{}lcccc@{}}
\toprule
                     &  & \textbf{Cam-CAN}        & \textbf{IXI} & \textbf{UK Biobank} \\ \midrule
\textbf{N}           &  & 652                     & 563          & 42,786                   \\ \midrule
\textbf{Age (years)} &  &                         &              &                     \\
Mean ± SD             &  & 54.3±18.6               & 48.6±16.5   & 64.0±7.7                   \\
Range                &  & 18 - 88                 & 20 - 86      &  44 - 82                   \\ \midrule
\textbf{Sex}         &  &                         &              &                     \\
Female/Male          &  & 330/332                 & 313/250      & 20,206/22,580                    \\ \midrule
\textbf{Race}        &  &                         &              &                     \\
White                &  & \multicolumn{1}{c}{---} & 451          &  41,417                   \\
Black                &  & \multicolumn{1}{c}{---} & 14           &  286                \\
Asian                &  & \multicolumn{1}{c}{---} & 50           &  454                   \\
Chinese              &  & \multicolumn{1}{c}{---} & 14           &  122                   \\
Other                &  & \multicolumn{1}{c}{---} & 34           &   507                  \\ \bottomrule
\label{demographic_table}
\end{tabular}
\end{table}

\paragraph{\textbf{Model}}

We adapted the conventional ResNet-34 model \cite{he2016deep} for age regression from 3D images. ResNet stands for Residual Network and is a type of CNN model with residual connections, a distinctive architecture designed to address the vanishing gradient problem during deep network training. We trained this model with whole preprocessed T1-weighted MRI images. The data was augmented through a composition of transformations, including random horizontal flip, contrast change, addition of Gaussian noise with random parameters and motion artifacts.

\subsection{Bias analysis}
    
We divided our statistical bias analysis into two parts, each focusing on a specific aspect. The first part aimed to assess bias in predictive performance, while the latter delved deeper into the model to examine biases in the generated features. To ensure a sufficient sample size for each subgroup, we considered the Chinese subjects to be part of the Asian group, and excluded all subjects with race classified as ``Other'' (which includes ``Mixed''). We then further divided each racial subgroup (``White'', ``Asian'' and ``Black'') into ``Female'' and ``Male'',  resulting in six test set subgroups. 
\paragraph{\textbf{Absolute performance assessment}} We calculated the absolute error of prediction, using it as the main performance metric. With the goal of comparing the performance between all subgroups, we then progressed by verifying the assumptions necessary to perform an Analysis of Variance (ANOVA), i.e. assumption of normality - through visual inspection of the absolute error distribution and Shapiro-Wilk tests - and the assumption of homogeneity of variances, through the Levene's test. The assumption of sample independence is met from the experimental design, as all subgroups are constituted by different subjects. Given that not all assumptions were met, we progressed by using the non-parametric Kruskal-Wallis test to compare the absolute error medians of all subgroups. Further pairwise comparisons were completed using the post-hoc Conover-Iman test. Since the Kruskal-Wallis test is the non-parametric equivalent of the one-way ANOVA, we conducted two Conover-Iman tests in order to take into consideration our two factors, race and biological sex.
Although the Kruskal-Wallis test can handle unbalanced data, when the differences are large its power is reduced, which may lead to inconsistent/intransitive results \cite{brunner2018rank}. In order to ensure the validity and consistency of our results, we balanced the data by randomly selecting a sample from each subgroup with equal sample size. The sample size chosen for each subgroup was 126, i.e. the size of the smallest group (Black female subjects). After calculating the mean absolute error for each subgroup sample, we repeated the random sampling ten times to estimate the standard deviation. We then repeated the statistical procedure, verifying ANOVA's assumptions, and upon rejection of normality following with the Kruskal-Wallis test and corresponding post-hoc Conover-Iman tests.\\

\paragraph{\textbf{Model features assessment}} Additionally, we assessed if the features generated by the model were biased using the framework for feature exploration proposed by Glocker et al \cite{glocker2023algorithmic}. This strategy consists of passing each test set scan through our model up to the penultimate layer, extracting its output features and subsequently inputting them to a principal component analysis (PCA) model in order to reduce their dimensionality. The PCA projections consist of a new set of dimensions (also called ``modes'') which capture the directions of the largest variation in the high-dimensional feature space. Given that our model was trained to predict age, it is expected that the strongest separation for samples in different age groups is seen in the first PCA modes. We plotted the distribution of samples in PCA space (first four modes) through kernel density estimation plots, split by the three demographic attributes of interest (age, biological sex, and race). Age was divided into five brackets to facilitate visual analysis. This was accompanied by two-sample Kolmogorov-Smirnov tests to compare the feature distributions of all possible pairwise combinations across race, sex, and age subgroups, in the first four modes of PCA. To decrease the number of statistical tests to be completed, for this step age was divided into only two brackets, [40-60] and [60-90]. To account for multiple testing and the consequent type 1 error inflation, the p-values were adjusted using the Benjamini-Yekutieli procedure. We considered statistical significance at a 95\% confidence level.

\section{Results}
\begin{figure}[h!]
\centering
\includegraphics[scale=0.4]{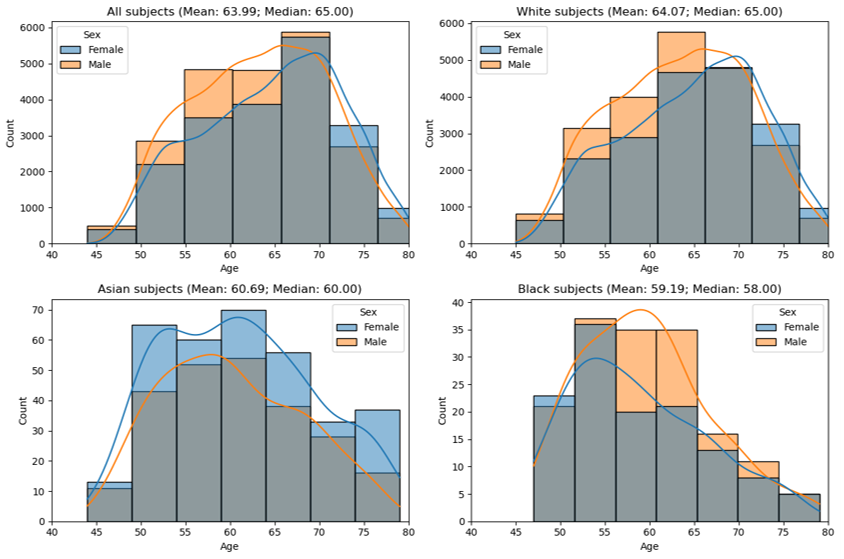}
\caption{Age distribution of all subjects in the test set and of each racial subgroup, separated by biological sex. Overlapping lines show the probability density curves.}
\label{fig:age dist all}
\end{figure}

Figure \ref{fig:age dist all} shows the age distribution for each subgroup, including all samples available in the test set. Within White subjects, we can observe a tendency of younger male and older female subjects, whereas for Black subjects, we find the opposite.

\begin{figure}[h!]
\centering
\includegraphics[scale=0.3]{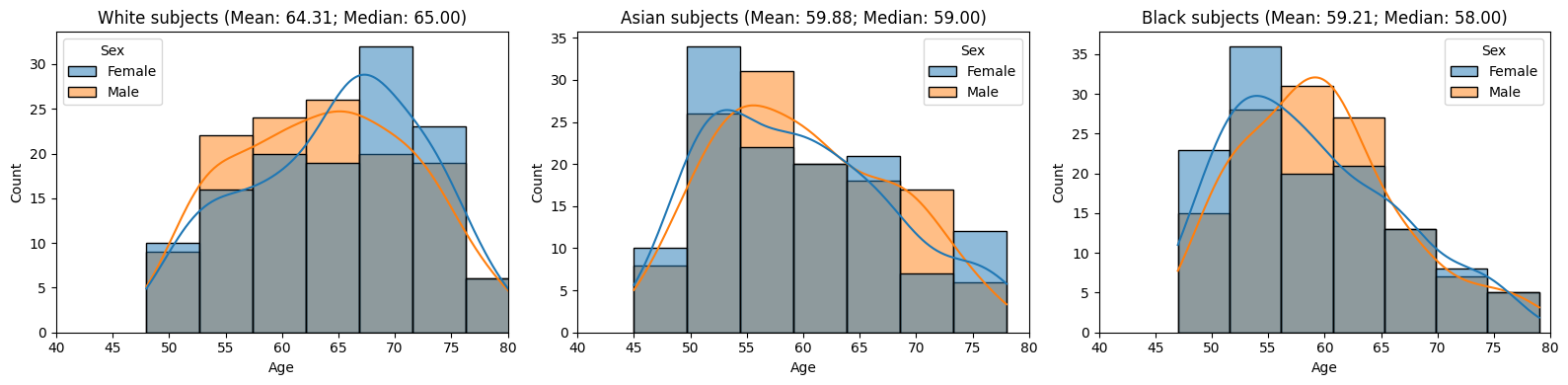}
\caption{Age distribution of a random sample (n=126) from each racial subgroup, separated by biological sex. Overlapping lines show the probability density curves.}
\label{fig:agedist_sample}
\end{figure}

\paragraph{\textbf{Absolute performance assessment}} Our first analysis involved conducting a Shapiro-Wilk test to evaluate whether the subgroup's prediction errors followed a normal distribution, and the Levene's test to verify whether all groups to be compared had equal variance. The resulting p-values from the six Shapiro-Wilk tests were below the defined significance level of 0.05. This indicates that we can confidently reject the null hypothesis that the population from which each sample is drawn follows a normal distribution. For visual confirmation, the distribution of the absolute error for each subgroup, along with the corresponding p-values from the Shapiro-Wilk tests are given in the Appendix (Figure \ref{fig:norm_all}). On the other hand, the Levene's test returned a p-value of \num{5.82e-52}, confirming that we have sufficient evidence to reject the null hypothesis and conclude that not all samples come from populations with equal variances. Upon the rejection of both assumptions for two-way ANOVA, we proceeded by conducting a Kruskal-Wallis test. The resulting p-value was \num{6.99e-116}, leading us to reject the null hypothesis, i.e. that the population medians are all equal. The resulting p-values from the two Conover-Iman tests conducted for further pairwise comparisons were as follows: White vs Asian: 0.022; White vs Black: 0.017; Black vs Asian: 0.0015; female vs male: \num{4.20e-118}. As suspected, the p-values from both the Kruskal-Wallis and the Conover-Iman test were notably low. The plot in Figure \ref{fig:agedist_sample} shows the age distribution of the random samples from each subgroup (n=126), taken to ensure the robustness of our statistical procedure. It reveals a similar pattern to that observed when examining all subjects in the test set, showing higher prevalence of younger White males and older White females, while the opposite is seen for Black subjects.\\

\begin{figure}[h!]
\centering
\includegraphics[scale=0.5]{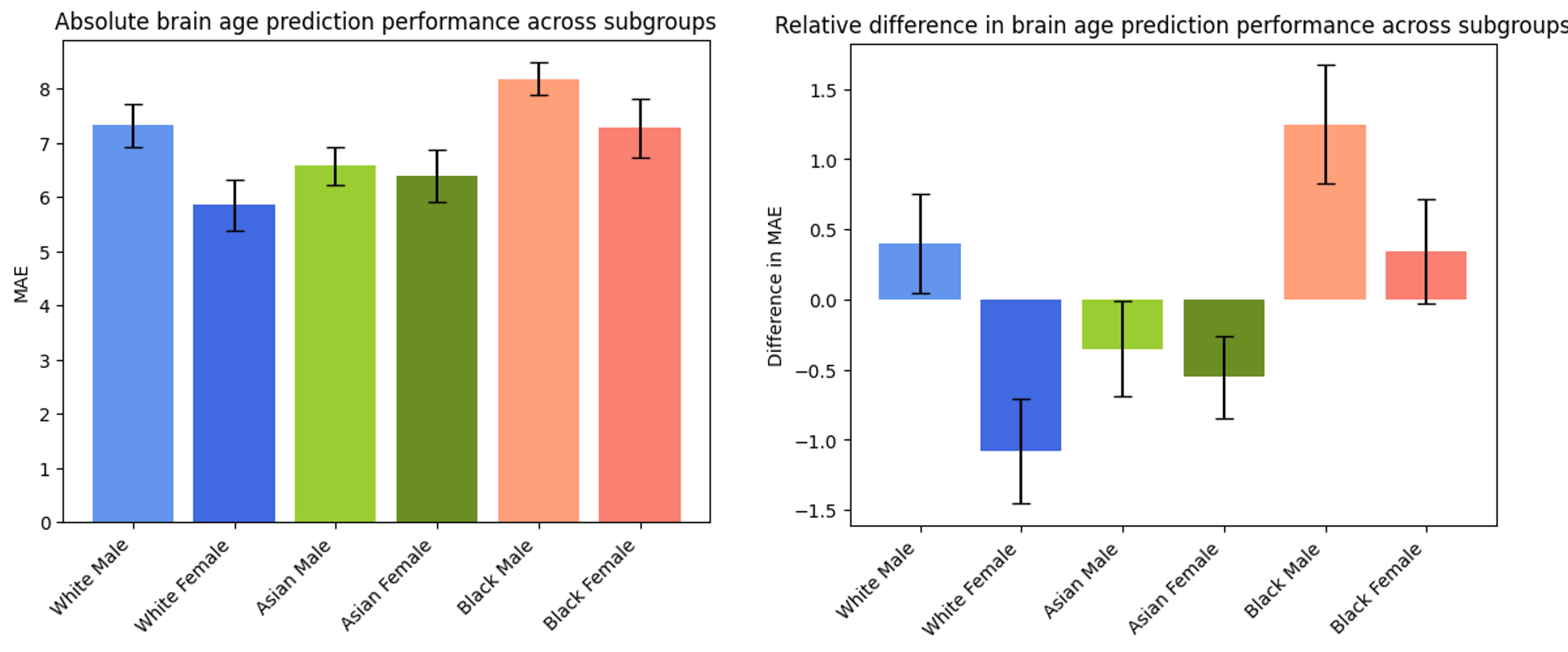}
\caption{Left: MAE, considering only a random sample of 126 subjects for each subgroup. Right: Relative difference in brain age prediction performance across patient subgroups. Difference calculated in relation to the average absolute error across all subjects (i.e. all random samples, with a total of 756 subjects). For both plots, the error bars were created by repeating the random sampling ten times and calculating the standard deviation.}
\label{fig:maeplot}
\end{figure}

The first plot in Figure \ref{fig:maeplot} shows the mean absolute error (MAE) for each subgroup sample, accompanied by its corresponding error bar, calculated from ten random samples of the same dimension. The adjacent plot illustrates the disparity in absolute error concerning the average absolute error across all subjects. Notably, we see a considerable under-performance of the model for Black male subjects. The model achieves its highest performance on White female and Asian subjects.\\

Similarly to the tests conducted with all test set subjects, the p-values resulting from the Shapiro-Wilk tests for normality testing indicated that we could reject the null hypothesis that the samples come from a normally-distributed population, with a significance level of 0.05. Contrarily, the Levene's test produced a p-value of 0.78, suggesting that we cannot reject the null hypothesis of equal variances across all samples. However, as not all ANOVA assumptions were met, we proceeded with the Kruskal-Wallis test, which yielded a p-value of 0.0015. With a p-value below our defined significance level (0.05), we reject the null hypothesis and have sufficient evidence to suggest that the differentiating factors among subgroups lead to statistically significant differences in the model's performance. The resulting p-values from the two post-hoc Conover-Iman tests were the following: White vs Asian: 0.76; White vs Black: 0.022; Black vs Asian: 0.013; female vs male: 0.008. These outcomes reveal statistically significant disparities in the model's performance between White and Black subjects and Black and Asian subjects, as well as between female and male subjects.\\

\begin{figure}[h!]
\centering
\includegraphics[scale=0.5]{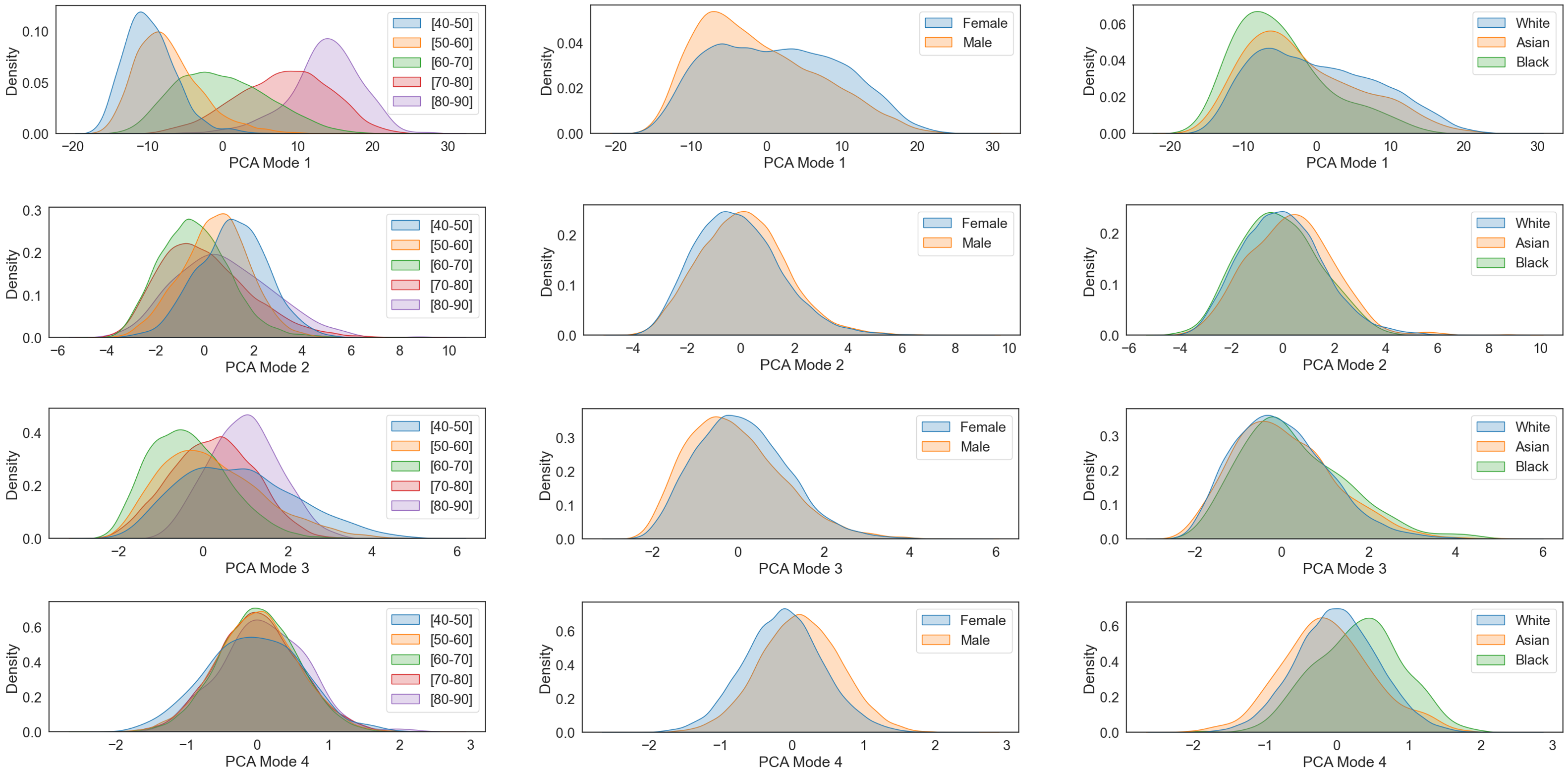}
\caption{Kernel density estimation plots depicting the density distribution of each age, race, and biological sex subgroup across the first four PCA modes of the feature space.}
\label{fig:pca_results}
\end{figure}

\paragraph{\textbf{Model features assessment}} Proceeding to the examination of bias in the model's features, the kernel density estimation plots presented in Figure \ref{fig:pca_results} show the density distribution of each age, race, and biological sex subgroup as generated by PCA in its first four modes. These plots include all subjects available in our test set. Here, we can infer that the PCA modes of primary interest are modes 1, 2 and 3, as they show the strongest separation between age groups, aligning with the model's training objective. Therefore, we are particularly interested in examining subgroup differences in these modes. It is nevertheless worth noting that in PCA mode 4 there is a clear separation between racial subgroups. However, these disparities might not be of primary concern as these features may not be informative for age prediction.\\

The adjusted p-values from the two-sample Kolmogorov-Smirnov tests, conducted to compare the distributions of each pair of subgroups, can be found in Table \ref{tab:stat_pca_results_all} of the Appendix. Similarly to the procedure described above, the tests in which all subjects available in the test set were used yielded notably low p-values and rendered almost all pairwise comparisons statistically significant, with only three exceptions: White vs Black in mode 2, Asian vs White in mode 3, and Black vs Asian in mode 3. The adjusted p-values from the new Kolmogorov-Smirnov tests, including only a equal-sized sample of each subgroup, are shown in Table \ref{tab:stat_pca_results}. When looking at the first three PCA modes and comparing racial subgroups, we find five of the nine pairwise comparisons between marginal distributions to be statistically significant. For biological sex, on the other hand, two of the three comparisons were statistically significant.

\begin{table}
    \centering
    \renewcommand{\arraystretch}{1.5}
    \setlength{\tabcolsep}{3pt}
    \caption{P-values resulting from two-sample Kolmogorov-Smirnov tests which compared marginal distributions from the pairs of subgroups indicated, across the first four PCA modes. Results including a random sample of each subgroup, all with equal size. The p-values were adjusted for multiple testing using the Benjamini-Yekutieli procedure. Significance level is set to 0.05. Statistically significant results are coloured with red.}
    \begin{tabular}{l *{5}{>{\columncolor{white}[0pt][\tabcolsep]}c}}
        \hline
        & Age 40-60/60-90 & Asian/White & Black/Asian & White/Black & Female/Male \\
        \hline
        PCA mode 1 & \cellcolor{Maroon!30}<0.0001 & \cellcolor{Maroon!30}0.0065 & \cellcolor{Maroon!30}0.011 & \cellcolor{Maroon!30}<0.0001 & \cellcolor{YellowGreen!30}0.39 \\
        PCA mode 2 & \cellcolor{Maroon!30}<0.0001 & \cellcolor{Maroon!30}0.011 & \cellcolor{Maroon!30}0.0065 & \cellcolor{YellowGreen!30}1 & \cellcolor{Maroon!30}0.006 \\
        PCA mode 3 & \cellcolor{Maroon!30}<0.0001 & \cellcolor{YellowGreen!30}0.52 & \cellcolor{YellowGreen!30}0.52 & \cellcolor{YellowGreen!30}0.13 & \cellcolor{Maroon!30}0.0037 \\
        PCA mode 4 & \cellcolor{YellowGreen!30}0.09 & \cellcolor{YellowGreen!30}0.25 & \cellcolor{Maroon!30}<0.0001 & \cellcolor{Maroon!30}<0.0001 & \cellcolor{Maroon!30}0.025 \\
        \hline
    \end{tabular}
    \label{tab:stat_pca_results}
\end{table}

\section{Discussion and conclusion}

In this study, we aimed to thoroughly investigate the potential race and biological sex bias in a model for brain age prediction from MRI predominantly trained on White subjects. The statistical tests conducted to evaluate the model's absolute performance reveal statistically significant differences for Black subjects, compared to both Asian and White subjects, as well as differences between male and female subjects. When looking back at the model's average performance per subgroup (Figure \ref{fig:maeplot}), we can observe the same evidence, concluding that the negatively affected groups are Black and male subjects. One possible explanation might be the fact that these are the two most underrepresented groups in our training set, which contained 582 male (versus 643 female) subjects, and only 14 Black (versus 451 White and 64 Asian) subjects. The imbalance in racial distribution for the training set only include the IXI dataset as race information was not available for the Cam-CAN dataset. Here, we assumed that Cam-CANis predominantly White. Additionally, the results of our statistical analysis over the model's feature inspection suggests that some of the features that encode information useful for age prediction, also allow for the separation of both racial and biological sex subgroups.\\

In practicality, recent research primarily focuses on assessing the correlation between brain age gap and neurological disorders/clinical risks. This gap represents the model's prediction error, which can be attributed to noise (model accuracy, data quality) and physiology. When evaluating the latter, it's crucial to distinguish disorder-related changes from inherent biological differences due to sex or ethnicity. This study reveals an average of 1 to 2-year statistically significant disparities among ethnicity subgroups. Depending on biomarker application, these deviations hold significance. For instance, Cole et al. (2018) \cite{cole2018brain} found a 6.1\% rise in mortality risk between ages 72-80 per extra predicted brain year. Accounting for ethnicity could be vital in such cases.\\

Lange et al. \cite{de2022mind} have previously reported that the metrics used to evaluate brain age prediction performance, including MAE, are significantly affected by discrepancies in the age range of the training and testing datasets. One limitation of our study is hence the limited age range of UK Biobank (44-82) - our test set - when compared to the broad range encompassed by the training set (18-88), which is desired for an age prediction model. As a consequence, we might observe a lower overall age prediction performance than the state-of-the-art. Nevertheless, given that our primary goal was to compare the model's performance across subgroups, and that the age range is similar across the random samples of each test subgroup (Figure \ref{fig:agedist_sample}), we can assume that this evaluation remains meaningful despite age range variations in training and test sets.\\

Another limitation of our study is the use of a single model type, one combination of datasets and a specific type of input features (T1-weighted MRI scans). However, we believe that our findings are relevant to further motivate a systematic bias assessment, including a diverse range of commonly employed models, such as other CNN models, ensembles, or simpler machine learning models like XGBoost, as these have been shown to have comparable performance to more complex DL models \cite{more2023brain}. Another interesting avenue for exploration would be to examine whether the similar biases persist when employing MRI-derived features, e.g. white and grey matter maps or volumes of subcortical structures.\\

Our results suggest that training brain age prediction on imbalanced data leads to significant differences in subgroup performance. We call for comprehensive bias assessment in other brain age prediction models, as these have emerged as important diagnostic and prognostic clinical tools.

\section{Acknowledgments}
B.G. is grateful for the support from the Royal Academy of Engineering as part of his Kheiron Medical Technologies/RAEng Research Chair in Safe Deployment of Medical Imaging AI. C.P gratefully reports financial support provided by UKRI London
Medical Imaging \& Artificial Intelligence Centre for Value Based
Healthcare.
\newpage
\printbibliography

\newpage
\appendix
\renewcommand{\thefigure}{\thesection.\arabic{figure}}
\setcounter{figure}{0}
\section{Appendix}

\begin{figure}[h!]
\centering
\includegraphics[scale=0.3]{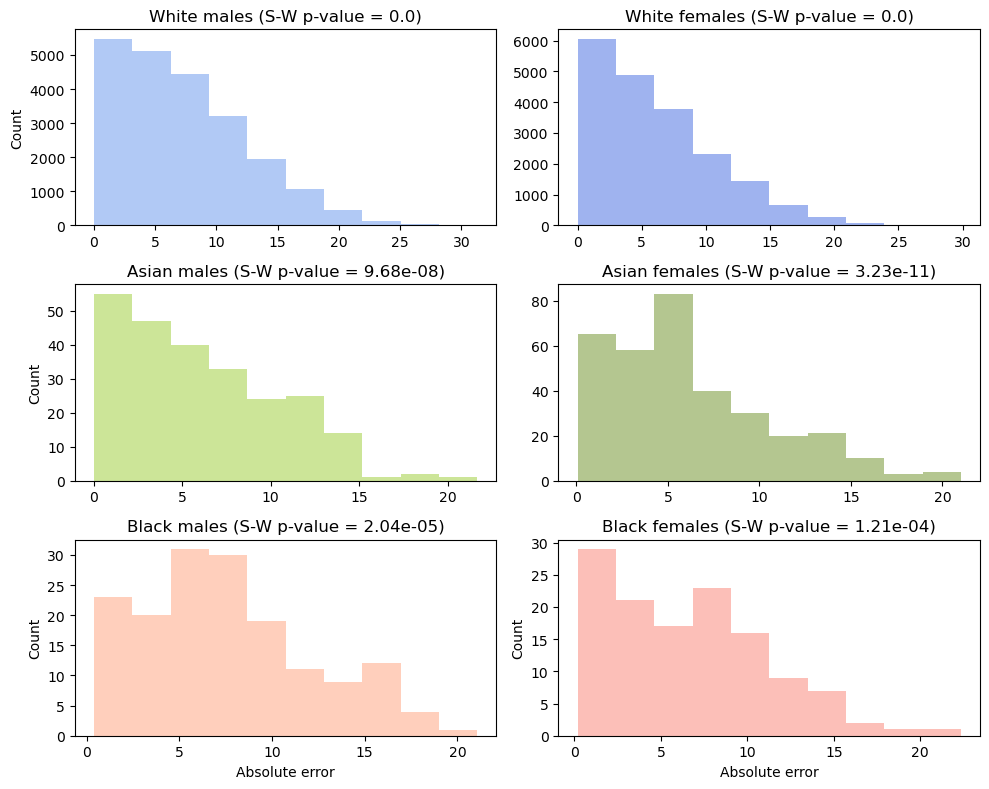}
\caption{Histograms showcasing the distribution of the absolute prediction error for each subgroup, including all subjects in test set.}
\label{fig:norm_all}
\end{figure}

\setcounter{table}{0}
\renewcommand{\thetable}{\thesection.\arabic{table}}

\begin{table}
    \centering
    \renewcommand{\arraystretch}{1.5}
    \setlength{\tabcolsep}{3pt}
    \caption{P-values resulting from two-sample Kolmogorov-Smirnov tests which compared marginal distributions from the pairs of subgroups indicated, across the first four PCA modes. Results including all samples available in the test set. The p-values were adjusted for multiple testing using the Benjamini-Yekutieli procedure. Significance level is set to 0.05. Statistically significant results are coloured with red.}
    \begin{tabular}{l *{5}{>{\columncolor{white}[0pt][\tabcolsep]}c}}
        \hline
        & Age 40-60/60-90 & Asian/White & Black/Asian & White/Black & Female/Male \\
        \hline
        PCA mode 1 & \cellcolor{Maroon!30}<0.0001 & \cellcolor{Maroon!30}<0.0001 & \cellcolor{Maroon!30}<0.0001 & \cellcolor{Maroon!30}<0.0001 & \cellcolor{Maroon!30}<0.0001 \\
        PCA mode 2 & \cellcolor{Maroon!30}<0.0001 & \cellcolor{Maroon!30}<0.0001 & \cellcolor{Maroon!30}0.0071 & \cellcolor{YellowGreen!30}1 & \cellcolor{Maroon!30}<0.0001 \\
        PCA mode 3 & \cellcolor{Maroon!30}<0.0001 & \cellcolor{YellowGreen!30}0.78 & \cellcolor{YellowGreen!30}0.44 & \cellcolor{Maroon!30}0.039 & \cellcolor{Maroon!30}<0.0001 \\
        PCA mode 4 & \cellcolor{Maroon!30}0.025 & \cellcolor{Maroon!30}<0.0001 & \cellcolor{Maroon!30}<0.0001 & \cellcolor{Maroon!30}<0.0001 & \cellcolor{Maroon!30}<0.0001 \\
        \hline
    \end{tabular}
    \label{tab:stat_pca_results_all}
\end{table}

\end{document}